\documentclass[11pt,a4paper]{article}
\usepackage{cite}
\usepackage{graphicx}
\usepackage{epsfig}
\newcommand {\kn}{{\mathrm{K}^-}}
\newcommand {\nutau}{\nu_\tau}

\newcommand {\Be}{{\cal{B}}_{\mathrm{e}}}
\newcommand {\Bm}{{\cal{B}}_{\mu}}
\newcommand {\Bp}{{\cal{B}}_{\pi}}
\newcommand {\Bk}{{\cal{B}}_{\mathrm{K}}}
\newcommand {\Bl}{{\cal{B}}_{\ell}}
\newcommand {\Bh}{{\cal{B}}_{\mathrm{h}}}
\newcommand {\mixingfactor}{1-\sin^2\theta_{\mathrm{L}}}
\newcommand {\GF}        {G_{\mathrm{F}}}

\newcommand {\MNUTCvalintA}     {42} 
\newcommand {\MNUTCvalintB}   {48} 
\newcommand {\FMIXCvalintA}    {0.014} 
\newcommand {\FMIXCvalintB}    {0.017} 
\renewcommand {\Be}{B_{\mathrm{e}}}
\renewcommand {\Bm}{B_{\mu}}
\renewcommand {\Bp}{B_{\pi}}
\renewcommand {\Bk}{B_{\mathrm{K}}}
\renewcommand {\Bl}{B_{\ell}}
\renewcommand {\Bh}{B_{\mathrm{h}}}
\newcommand {\Bi}{B_i}
\begin{document}

\begin{titlepage}
\addtocounter{page}{-1}      

\vspace*{-1em}  

\hspace*{\fill}{\Large\bf{L3 Note 2196}}

\vspace*{2em}  

{\Large\bf                                                           
\begin{center}          
{\LARGE\bf 
Numerical Construction of Likelihood Distributions \\[2mm]
and the Propagation of Errors}
\end{center} }
\vspace*{2em}
{\normalsize
\begin{center}
{\Large \bf J. Swain$^{\dagger,1}$ and L. Taylor$^{\dagger,2}$}
\end{center} }
\vspace*{1.0em}
\begin{center}  
{\Large\today} \\[1mm]
\end{center}  

\vfill
\begin{center}
\parbox{0.75\textwidth}{%
\begin{center}
{\large\bf Abstract}\\[2mm]
\end{center}       

\noindent

The standard method for the propagation of errors, based 
on a Taylor series expansion, is approximate and frequently 
inadequate for realistic problems.
A simple and generic technique is described in 
which the likelihood is constructed numerically, thereby greatly 
facilitating the propagation of errors. 
}
\end{center}       

\vfill

\begin{center}  
{\large\em{To appear in Nucl. Instrum. Meth.}} \\[1mm]
\end{center}  

\vfill

{\small
\begin{tabular}{ll}
$^\dagger$  & Department of Physics, 
              Northeastern University, 
              360 Huntington Avenue, 
              Boston, MA 02115, USA \\
1           & Electronic mail: John.Swain@cern.ch    \\
2           & Electronic mail: Lucas.Taylor@cern.ch  \\
\end{tabular}
}
\end{titlepage}
 
\pagebreak
 
\section{Introduction}
Traditionally, errors on derived quantities have been determined analytically
by performing a Taylor series expansion about the central values.
This method is adequate for many simple problems but often fails 
in more realistic situations for a number of different reasons, 
as described in section~\ref{sec:standard}.

In section~\ref{sec:numerical} we describe a rigorous and generally 
applicable numerical technique for the propagation of errors,
based on the numerical construction of the likelihood for the 
derived quantity using pseudo-random numbers.
Sections~\ref{sec:wmass} and \ref{sec:nutaumass} illustrate the
use of this technique with practical examples from High Energy Physics 
analyses. 

While the method proposed has almost certainly been used by others, 
it has received little attention in the literature. 
This is perhaps due to the fact that only in the last few years has 
it been practical to generate large numbers of pseudo-random numbers in 
order to solve a problem which was traditionally done by hand. 

\section{Analytic Method for the Propagation of Errors\label{sec:standard}}

In general, physical quantities are not known with infinite precision 
but are described in terms of likelihood distributions (also called probability
or error distributions).
To be specific, consider a quantity $f$ which depends on some quantities
$x_i$ $(i=1,\ldots,n)$ in a known way.
The uncertainty on $f$, which depends on the uncertainties $\delta x_i$ on 
the measured values $x_i^0$, is
usually derived by the standard method for the propagation of errors 
which relies on a Taylor series expansion of $f$:
\begin{equation}
f(x_i^0 + \delta x_i) = f(x_i^0) + 
                      \sum_{i=1}^{n}
                      \delta x_i
                      \left.
                      \frac{\partial f}
                           {\partial x_i}
                      \right|_{x_i=x_i^0} 
                       + \cdots  
\label{equ:taylor}
\end{equation} 
where the ellipsis denotes higher order terms in $\delta x_i$.
These are neglected to yield the familiar expression for the variance of $f$: 
\begin{eqnarray}
\sigma_f^2  =    \sum_{ij} 
                 \sigma_{ij}
                 \left.
                 \left( \frac {\partial f}
                              {\partial x_i}
                        \frac {\partial f}
                              {\partial x_j}
                 \right)
                 \right|_{\begin{array}{l}
                            _{x_i = x_i^0;} \\[-0.3ex] 
                            _{x_j = x_j^0}
                          \end{array}}                   
\label{equ:varf}
\end{eqnarray}
where $\sigma_{ij}$ denotes the error matrix.
For the results of this method to be valid, a number of 
requirements should be satisfied.
\subsubsection*{Requirement 1: The likelihood distributions of $x_i$ are Gaussian.}
The standard method for the propagation of errors implicitly assumes that
the errors on the $x_i$ are Gaussian.
In general, however, a measured quantity is described by an 
arbitrarily shaped likelihood distribution.
Statistical and systematic errors on quantities $x_i$ may be according 
to other common likelihood distributions (Poisson, binomial, {\em{etc.}}),
or another function which may itself depend on  
measured quantities with associated errors. 
\subsubsection*{Requirement 2: The likelihood distribution of $f$ is Gaussian.}
The standard method for propagation of errors yields a single number
$\sigma_f$ which is interpreted as the RMS of the Gaussian likelihood
distribution of $f$.
It is often the case that the likelihood distribution of $f$ is 
not Gaussian, nor even symmetric, 
even if all the errors on $x_i$ are Gaussian.
In such cases the interpretation of $\sigma_f$ is unclear.
A familiar example occurs in tracking detectors where 
the inverse transverse momentum, $1/p_{\mathrm{T}}$, 
has a Gaussian error whereas the derived quantity,
$f(1/p_{\mathrm{T}}) = p_{\mathrm{T}}$, does not 
and is manifestly asymmetric.
\subsubsection*{Requirement 3: The required derivatives are calculable.}
The evaluation of the derivatives, ${\partial f}/{\partial x_i}$, may prove to 
be difficult in practice if $f$ is a complicated function of many parameters.
If the dependence of $f$ on $x_i$ were determined numerically, for example by
running a physics Monte Carlo program, it may be difficult or impossible
to determine the required derivatives.
Moreover, the derivatives may themselves have significant uncertainties
due to limited Monte Carlo statistics and systematic errors from
uncertainties in the values of physical parameters used by the 
Monte Carlo program.  
\subsubsection*{Requirement 4: Higher terms of the Taylor expansion are negligible.}
It is assumed that quadratic and even higher order terms in 
the Taylor expansion of Eq.~\ref{equ:taylor} are negligible.
This approximation is not always valid, for example if the first derivatives 
of $f$ at $x_i$ are zero or small compared to higher derivatives 
or if the errors are not small relative to the $x_i$.
%

\section{Numerical Method for the Propagation of Errors\label{sec:numerical}}
If one or more of the above conditions is not satisfied, we 
advocate the use of a numerical method for the propagation of errors,
as described below.
Even if all the conditions are satisfied in a given situation, the 
numerical method provides a valuable cross-check.
Moreover, if the situation becomes progressively more complicated 
as more sources of error are identified, it is easy to extend the numerical method,
whereas the standard method may ultimately fail one of the above assumptions. 

The numerical method is based on the familiar statistical 
concept of performing many hypothetical measurements of 
the same quantity and determining the error on that quantity
from the spread of the values.
It may be summarised as follows:
\begin{quotation}
{\noindent\it 
Consider a function $f(x_i)$ where the independent variables $x_i$
have uncertainties which are described by likelihood functions
$L({x_i})$.
The corresponding likelihood distribution $L(f)$ may be 
described with arbitrary precision 
by a sufficiently large set of values, $\{f(x_i^\prime)\}$, where
the input set of values, $\{x_i^\prime\}$,
are chosen randomly according to the $L({x_i})$.  
}
\end{quotation}
While the standard method for the propagation of errors based on the 
Taylor expansion of Eq.~\ref{equ:taylor} assumes a single $\delta x_i$,
which is equated to the standard deviation of a Gaussian,
the numerical method maps the full spectrum of errors $\delta x_i$
and therefore does not require it to be Gaussian.
Moreover, no higher order terms are neglected.

Practically one implements the numerical method as follows:
\begin{enumerate}
\item Define the likelihood functions, $L({x_i})$, 
      for the independent variables, $x_i$. \\
      For example, in the case of uncorrelated Gaussian errors one 
      would specify a set of central values and standard deviations.
      For correlated Gaussian errors one would specify a set of central values
      and a covariance matrix.  
      In a more general case one might define the likelihood as a multi-dimensional
      function represented by a smooth parametrisation, a binned multi-dimensional
      histogram, or as a set of discrete values derived, for example, from a 
      Monte Carlo program.
\item Repeat the following steps, (a) and (b), $n$ times, where $n$ is a large number:
      \begin{enumerate}
         \item Choose a set of values $\{x_i^\prime\}$ randomly according to
               the likelihood functions, $L({x_i})$. \\
               For example, in the case of uncorrelated Gaussian errors one 
               might use the CERNLIB\cite{WWW_CERNLIB} Fortran subroutine {\tt{RANNOR}}.
               For correlated Gaussian errors one could use the NAGLIB\cite{WWW_NAGLIB}
               Fortran subroutine {\tt{G05EAF}} or a suitably transformed set of 
               uncorrelated random \mbox{numbers\cite{SPAAN_RANDOM}.}
         \item Evaluate the function $f(x_i^\prime)$ and store the result, for example
               in a histogram or an array.
      \end{enumerate}
\item Normalise the integral of $L(f)$ to unity. 
\item Ensure that $n$ is sufficiently large, 
      for example by verifying that $L(f)$ does not change significantly
      for different starting random number seeds.
\end{enumerate}

Care should be taken to avoid numerical problems associated with the
computing hardware and software \mbox{used\cite[page 659]{JAMES91B}.}
For example, if $n$ is large then the accumulation of $f(x_i^\prime)$
values in a histogram may be subject to numerical imprecisions, or a 
poorly designed random number generator may start to show periodic behaviour.
The optimisers of some compilers may even treat multiple invocations of a 
random number function, for example in a Fortran {\tt{DO}} loop, 
as a single invocation and move it outside the loop thereby producing erroneous results.
Caveats such as these, however, apply to most software and can be 
avoided easily by minimal precautions such as the use of extended precision variables 
and low levels of compiler optimisation.
\section{Example: Measurement of the W Boson Mass\label{sec:wmass}} 
W mass measurements from hadron colliders\cite{PIC97}
are typically based on fits to distributions of transverse mass, 
$m_T = {\sqrt{2 p_T^\ell  p_T^\nu (1 - \cos \phi^{\ell\nu})}}$,
where
$p_T^\ell$ and $p_T^\nu$ denote the transverse momenta of the 
charged lepton (e or $\mu$) and the neutrino respectively
coming from the leptonic W decay,
and $\phi^{\ell\nu}$ denotes the angle between the charged lepton
and the neutrino in the transverse plane.
The charged lepton is typically measured in a tracking chamber with 
an error which is approximately Gaussian in $1/p_T^\ell$ while
$p_T^\nu$, which is inferred from the missing energy,
has an approximately Gaussian error. 
The opening angle error is typically relatively small.

Consider a single typical event with 
$p_T^\ell=40$\,GeV, 
$p_T^\nu=35$\,GeV,
$\phi^{\ell\nu}=2$\,rad,
which corresponds to $m_T = 63.0$\,GeV.
Assuming that the errors on $p_T^\ell$ and $p_T^\nu$ are uncorrelated, 
then the standard method for propagation of errors using Eq.~\ref{equ:varf}
yields
\begin{eqnarray}
\frac{\sigma_{m_T}^2}{m_T^2} 
& = & \left(\frac{1}{2 p_T^\ell}\right)^2  \sigma_{p_T^\ell}^2 +
      \left(\frac{1}{2 p_T^\nu}\right)^2   \sigma_{p_T^\nu}^2 \\
& = & \left(\frac{p_T^\ell}{2}\right)^2    \sigma_{1/p_T^\ell}^2 +
      \left(\frac{1}{2 p_T^\nu}\right)^2   \sigma_{p_T^\nu}^2 \\
\end{eqnarray}
For example, given
$\sigma_{1/p_T^\ell} = 0.005\,{\mathrm{GeV}}^{-1}$ and
$\sigma_{p_T^\nu}/p_T^\nu = 10\%$ yields 
$\sigma_{m_T} = 7.0$\,GeV.

\begin{figure}[tbp!]
  \begin{center}
     \epsfig{file=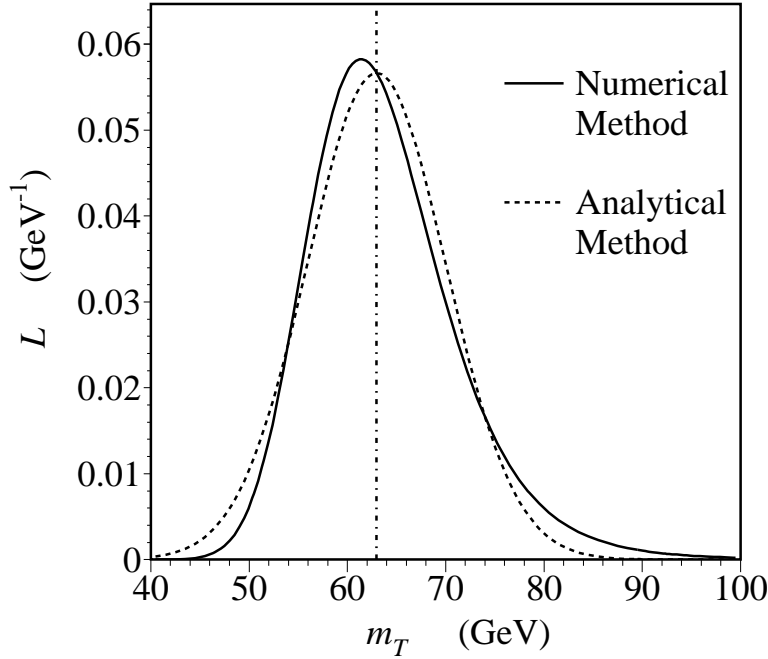,width=0.6\textwidth,clip=}
  \end{center}
  \caption{Example likelihood distribution of transverse mass showing the difference between
            the numerical and analytic approaches described in the text.
            Both curves are normalised to unit area.}\label{fig:mt}
\end{figure}

The likelihood distribution of $m_T$ is not in reality a Gaussian.
Fig.~\ref{fig:mt} (solid line) shows the $m_T$ likelihood distribution 
obtained by propagating the 
uncertainties on $p_T^\ell$ and $p_T^\nu$ numerically.
A clear asymmetry is seen, in contrast to the naive method for propagation of errors
which, by construction, yields a Gaussian likelihood distribution (dashed line). 
The shift in the peak is $\sim 2$\,GeV which is significant compared 
to typical errors on the W mass of $O(0.1)$\,GeV.

\section{Example: $\tau$ neutrino mass and mixing\label{sec:nutaumass}}
Constraints on the tau neutrino mass and its mixing with a hypothetical 
fourth lepton generation have been derived\cite{MNUTAU} by considering 
the dependence of tau branching fractions on the mass $m_{\nu_\tau}$
and the Cabibbo-like mixing angle $\theta_L$ 
(or more naturally $\sin^2\theta_L$).  
The theoretical predictions are compared 
with the experimental measurements for the following decays%
\footnote{Henceforth we denote the branching ratios for these processes as
          $\Be, \Bm, \Bp, \Bk$ respectively;
          $\Bl$ denotes either $\Be$ or $\Bm$ while $\Bh$ denotes either $\Bp$ or $\Bk$.}:
$\tau^-\rightarrow\mathrm{e}^-\bar{\nu}_{\mathrm{e}}\nutau$,
$\tau^-\rightarrow\mu^-\bar{\nu}_\mu\nutau$,
$\tau^-\rightarrow\pi^-\nutau$, and
$\tau^-\rightarrow\mathrm{K}^-\nutau$. 

The experimental measurements of the branching ratios
$B_i^{\mathrm{expt}}$ ($i={\mathrm{e, \mu, \pi, K}}$) are uncorrelated 
and have Gaussian errors.
If the theoretical predictions, $\Bi^{\mathrm{theory}}$, were perfectly known,
then the likelihood for a particular choice of $m_{\nu_\tau}$ and $\sin^2\theta_L$ would
be\footnote{The CLEO measurement of the $\tau$ mass was used to further constrain $m_{\nu_\tau}$.
            From an analysis of 
            $\tau^+\tau^-$ 
            $\rightarrow$ 
            $(\pi^+n\pi^0\bar{\nu}_\tau)$
            $(\pi^-m\pi^0\nu_\tau)$ 
            events (with $n\leq2, m\leq2, 1\leq n+m\leq3$), CLEO determined the $\tau$ mass to be 
            $m_\tau = (1777.8 \pm 0.7 \pm 1.7) + [m_{\nu_\tau}(MeV)]^2/1400 MeV$\cite{CLEOWEINSTEIN}.
            The likelihood for the CLEO and BES measurements to agree, as a function of 
            $m_{\nu_\tau}$ is included in the global likelihood.
            This does not affect the conclusions of this section but merely reinforces them.} 
\begin{equation}
L (m_{\nu_\tau},\sin^2\theta_L | \Be^{\mathrm{expt}},\Bm^{\mathrm{expt}},\Bp^{\mathrm{expt}},\Bk^{\mathrm{expt}})
= \prod_i \frac{1}{\sqrt{2\pi}\sigma_i} 
          {\mathrm{exp}}\left(
          -\left(\Bi^{\mathrm{theory}}-\Bi^{\mathrm{expt}}\right)^2 \left/ 2\sigma_i^2 \right. 
          \right),
\label{equ:likelihood}
\end{equation}
where the $\sigma_i$ are the errors on the $B_i^{\mathrm{expt}}$.
The predictions for $\Bi^{\mathrm{theory}}$, however, depend in turn on experimentally
measured quantities with errors. 

The theoretical predictions for the branching fractions $\Bl$ for the 
decays $\tau^-\rightarrow\ell^-\bar{\nu}_{\ell}\nutau$, 
with $\ell={\mathrm{e}}/\mu$, are given by\cite{MNUTAU}:
\begin{eqnarray}
  \Bl^{\mathrm{theory}}  
              & = & \left(\frac {\GF^2 m_\tau^5}{192\pi^3}\right)\tau_\tau (\mixingfactor)                                            \nonumber \\
              &   & \times  \left[ 1 -  8x - 12 x^2{\mathrm{ln}}x + 8 x^3 - x^4 - 8y(1-x)^3+\cdots \right]      \nonumber \\
              &   & \times     \left[ \left( 1 - \frac{\alpha(m_\tau)}{2\pi} \left( \pi^2 - \frac{25}{4} \right) \right)
                    \left( 1 + \frac{3}{5} \frac{m_\tau^2}{m_{\mathrm{W}}^2} + \cdots \right)\right]                    
\label{equ:blept}
\end{eqnarray}
where $\GF$ is the Fermi constant, 
$m_\tau$ and $\tau_\tau$ are the $\tau$ mass and lifetime, 
$x=m_\ell^2/m_\tau^2$,
$m_\ell$ is the charged lepton mass,  
$y=m_{\nu_\tau}^2 / m_\tau^2$,
$m_{\mathrm{W}}$ is the W mass,
$\alpha(m_\tau)$ is the renormalised fine-structure constant at the $\tau$ mass scale,
and each ellipsis denotes neglected higher order terms.
The first term in brackets allows for phase-space
while the second term in brackets allows for 
radiative corrections\cite{BERMAN58A,KINOSHITA59A,SIRLIN78A,MARCIANO88A}.
Similarly, the branching fractions for the decays $\tau^-\rightarrow\mathrm{h}^-\nutau$, 
with $\mathrm{h}=\pi/\mathrm{K}$, are given by
\begin{eqnarray}  
  \Bh^{\mathrm{theory}}  
             &  = &    \left(\frac {\GF^2 m_\tau^3 } {16\pi}\right)\tau_\tau f_{\mathrm{h}}^2 |V_{\alpha\beta}|^2 (\mixingfactor) \nonumber \\
             &    &    \times \left[ (1 - x)^2 \left ( 1 - y \left( \frac{2+x-y} {1-x} \right) \right) 
                                        \sqrt{ 1 - y \left( \frac{2+2x-y} {(1-x)^2} \right)}                       
                       \right]                                                                             \nonumber \\         
             &    &    \times \left[
                       1 + \frac{2\alpha}{\pi} \mathrm{ln} \left( \frac {m_{\mathrm{Z}}} {m_\tau} \right)+\cdots 
                       \right]                                                                             
         \label{equ:bhad}
\end{eqnarray}
where 
$x=m_{\mathrm{h}}^2 / m_\tau^2$,
$m_{\mathrm{h}}$ is the hadron mass, 
$y = m_{\nu_\tau}^2 / m_\tau^2$,  
$f_{\mathrm{h}}$ are the hadronic form factors, 
and $V_{\alpha\beta}$ are the CKM matrix elements, 
$V_{\mathrm{ud}}$ and $V_{\mathrm{us}}$,
for $\pi^-$ and $\kn$ respectively.
The ellipsis represents terms, estimated to be ${\cal{O}}(\pm 0.01)$\cite{MARCIANO92A},        
which are neither explicitly treated nor implicitly absorbed into $\GF$,
$f_\pi |V_{\mathrm{ud}}|$, or $f_{\mathrm{K}} |V_{\mathrm{us}}|$.

The uncertainties on the $\Bi^{\mathrm{theory}}$ depend 
on the errors on the values of: 
$\GF$, 
$\tau_\tau$,
$m_{\mathrm{e}}$, 
$m_\mu$, 
$m_\pi$, 
$m_{\mathrm{K}}$,   
$m_{\mathrm{W}}$, and         
$m_{\mathrm{Z}}$\cite{PDG96};
$m_\tau$ from the BES measurement at threshold\cite{MTAUBESNEW};        
$f_\pi |V_{\mathrm{ud}}|$ and 
$f_{\mathrm{K}} |V_{\mathrm{us}}|$\cite[and references therein]{MARCIANO92A};
and the estimated theoretical uncertainty due to (neglected) 
higher order radiative corrections.

If the standard method for the propagation of errors were to be applied, one 
would calculate the theoretical errors according to Eq.~\ref{equ:varf}, by
differentiation of Eqs.~\ref{equ:blept} and \ref{equ:bhad}, and add
them in quadrature to the experimental errors on the $B_i$ to obtain $\sigma_i$.
This approach is problematic for a number of reasons:
\begin{itemize}
\item the input errors are not necessarily Gaussian (for example the theoretical
      uncertainty on the neglected higher order terms); 
\item the uncertainty on $B_i^{\mathrm{theory}}$ is non-Gaussian,
      as may be seen immediately from just the $m_\tau^5$ and $m_\tau^3$ dependences 
      of $\Bl^{\mathrm{theory}}$ and $\Bh^{\mathrm{theory}}$ respectively;
\item many rather lengthy derivative calculations are required;
\item there is no {\em{a priori}} guarantee that the neglect of higher order terms 
      in the Taylor expansion is a reasonable approximation;
\item the four $\Bi^{\mathrm{theory}}$ predictions depend on many common input 
      parameters such that the four $\sigma_i$ cannot be treated as independent errors.
\end{itemize}

\begin{figure}[!tbp]
\begin{center}
    \mbox{\epsfig{file=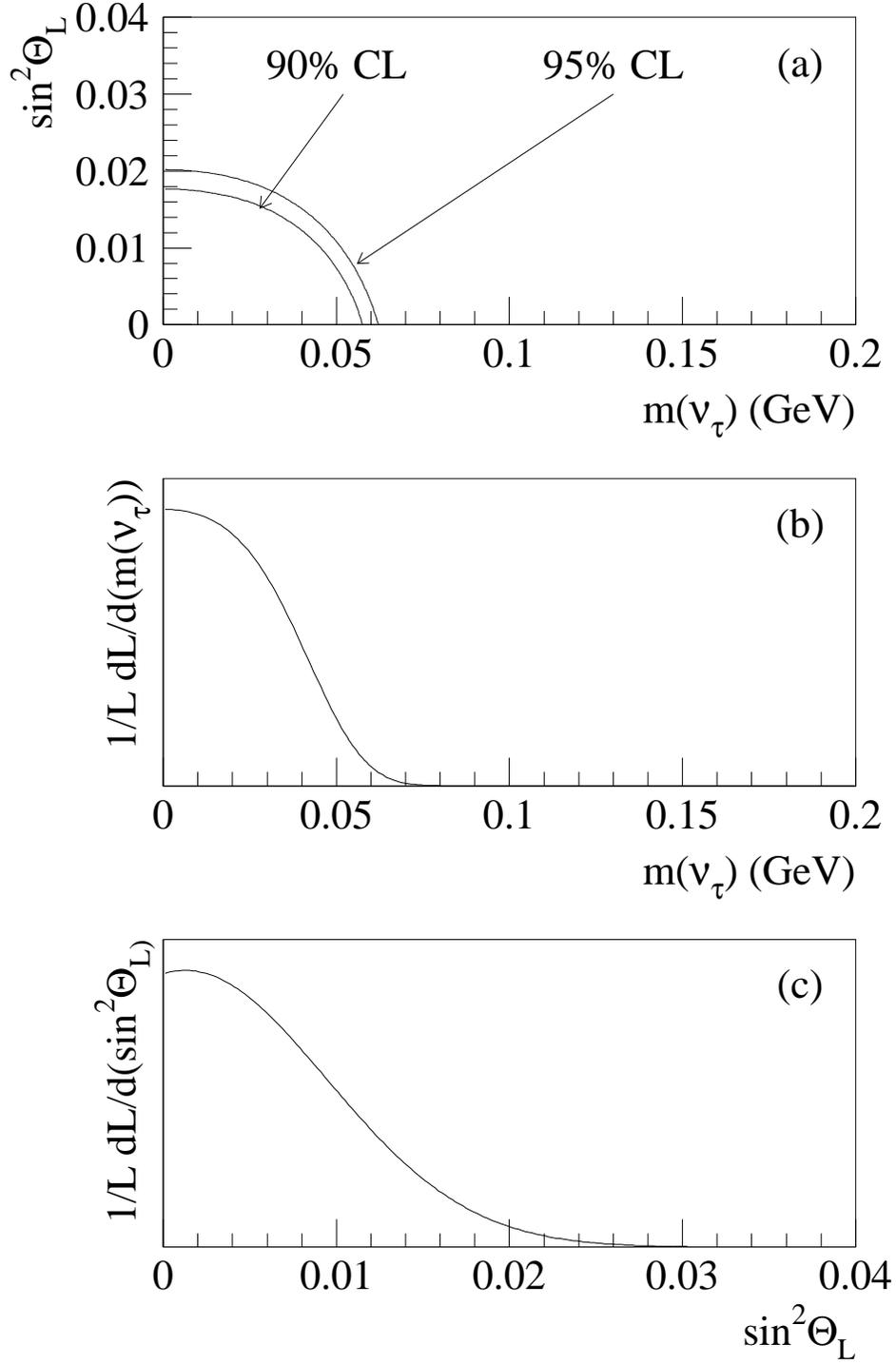,height=0.85\textheight,clip=}}
\end{center}
  \caption{Likelihood distributions for all $\tau$ decay channels combined, for 
(a) $\sin^2\theta_{\mathrm{L}}$ {\em{vs.}} $m_{\nu_\tau}$,
(b) $m_{\nu_\tau}$, integrated over $\sin^2\theta_{\mathrm{L}}$, and 
(c) $\sin^2\theta_{\mathrm{L}}$, integrated over $m_{\nu_\tau}$.
}\label{fig:contour}
\end{figure}

The numerical procedure described in section~\ref{sec:numerical} 
avoids all of these problems.
A large ensemble of values of $\Bi^{\mathrm{theory}}$ is 
created by choosing values for $\GF$, $\tau_\tau$, etc. according
to their errors and then evaluating   
$\Be^{\mathrm{theory}}$,
$\Bm^{\mathrm{theory}}$,
$\Bp^{\mathrm{theory}}$, and
$\Bk^{\mathrm{theory}}$
according to Eqs.~\ref{equ:blept} and \ref{equ:bhad}.
The likelihood is calculated according to 
Eq.~\ref{equ:likelihood} with $\sigma_i$ taken to be the error 
on $\Bi^{\mathrm{expt.}}$ only. 
The full likelihood, allowing for the errors on 
$\Bi^{\mathrm{expt.}}$, 
$\Bi^{\mathrm{theory}}$, 
and all correlations is then obtained from the 
normalised sum of the likelihoods for the full ensemble.
To be specific, the following steps are carried out:
\begin{enumerate}
\item define the likelihood functions, $L({x_i})$, 
      for the independent variables, $x_i$;
\item create a 2D histogram of $m_{\nu_\tau}$ vs. $\sin^2\theta_{\mathrm{L}}$;
\item repeat the following steps, (a) and (b), $n$ times:
      \begin{enumerate}
         \item choose a set of values $\{x_i^\prime\}$ randomly according to
               the likelihood functions, $L({x_i})$.
         \item for each bin in the histogram choose $m_{\nu_\tau}$ and 
               $\sin^2\theta_{\mathrm{L}}$ at the centre of the bin and then 
               evaluate 
               $L (m_{\nu_\tau},\sin^2\theta_L  | \Be^{\mathrm{expt}},\Bm^{\mathrm{expt}},
               \Bp^{\mathrm{expt}},\Bk^{\mathrm{expt}})$,
               according to Eqs. 
               \ref{equ:likelihood},
               \ref{equ:blept}, and
               \ref{equ:bhad},
               and add the value of $L$ to the contents of the bin; 
      \end{enumerate}
\item normalise the histogram to unity to obtain $L (m_{\nu_\tau},\sin^2\theta_L)$.
\end{enumerate}
Figure \ref{fig:contour}(a)\ shows the contours of the two dimensional 
likelihood distribution, $L (m_{\nu_\tau},\sin^2\theta_L)$ which
correspond to the 90\% and 95\% confidence levels.
No evidence is seen for a non-zero neutrino mass, nor for mixing.
By integration of the two dimensional likelihood over all values of 
$\sin^2\theta_{\mathrm{L}}$ we obtain the one-dimensional 
likelihood for $m_{\nu_\tau}$, 
as shown by the solid line of figure \ref{fig:contour}(b), which yields upper limits of 
$m_{\nu_\tau}<\MNUTCvalintA(\MNUTCvalintB)$\,MeV at the 90(95)\% confidence levels.
The solid line of figure \ref{fig:contour}(c) shows the one-dimensional likelihood distribution for 
$\sin^2\theta_{\mathrm{L}}$, integrated over all values of 
$m_{\nu_\tau}$, from which we derive the upper limits:
$\sin^2\theta_{\mathrm{L}}<\FMIXCvalintA(\FMIXCvalintB)$ at the 90(95)\% confidence levels.

\section{Conclusions}

The standard method for the propagation of errors, based 
on a Taylor series expansion, is approximate and frequently 
inadequate for realistic problems.
In particular, it assumes that the errors on the independent quantities 
are Gaussian, that the error on the derived quantity is Gaussian,
that the required derivatives are calculable, and that
higher order terms in the Taylor expansion are negligible.

A numerical method for the propagation of errors is described which
makes no such assumptions, provides exact results with arbitrary
precision, and is straightforward to implement even for complicated problems.

Realistic examples illustrating this numerical technique are described.
The interpretation of constraints on neutrino masses, 
with either a {\em{Classical}} or a {\em{Bayesian}} approach\cite{EADIE71A},  
has received much attention in the literature\cite{PDG96,JAMES91A}.
In the example described herein, 
a flat Bayesian prior distribution has been implicitly assumed 
by sampling $m_{\nu_\tau}$ and $\sin^2\theta_{\mathrm{L}}$ from a 
histogram with uniform bins.
Similarly, negative neutrino masses are excluded by the choice of the 
histogram range.
While such choices are a matter of discussion, it should be emphasised 
that they are in no way imposed by the use of the numerical algorithm 
for the propagation of errors, which is in fact generally applicable.

\section*{Acknowledgements}
We would like to acknowledge many colleagues in the ARGUS and L3 
collaborations for valuable discussions, in particular Bernhard Spaan. 
This work was supported by the US National Science Foundation.


%

\begin{thebibliography}{99}
\bibitem{SPAAN_RANDOM}
{As B. Spaan has pointed out, it is straightforward to generate a set of random
  values $\vec{x}^\prime$ according to an arbitrary covariance matrix $V$.
  First determine the eigenvalues $\lambda_i$ of the covariance matrix and the
  corresponding eigenvectors $\vec{x}_i$. Construct a rotation matrix $R$ from
  the eigenvectors where each column corresponds to an eigenvector. Then,
  generate a vector $\vec{r}$ with $i$ uncorrelated elements each chosen
  randomly from the Gaussian distribution: $\vec{r}_i =
  \left[{\mathrm{exp}}\left(-{x^2}/{2\lambda_i}\right)\right] /
  \sqrt{2\pi\lambda_i}$. Obtain the set of correlated random numbers
  $\vec{x}^\prime$ by rotation: $\vec{x}^\prime = R\cdot\vec{r}$.}
\end{thebibliography}

\end{document}